\documentstyle[aps,prl,floats,graphicx,epsfig]{revtex}
\begin{document}
\draft
\twocolumn[\hsize\textwidth\columnwidth\hsize\csname @twocolumnfalse\endcsname

%
\title{Superfluid $^3$He, Particle Physics and Cosmology}
\author{G.E. Volovik  }

\address{  Low Temperature Laboratory, Helsinki
University of Technology, Box 2200, FIN-02015 HUT, Finland\\
and\\ Landau Institute for Theoretical Physics, Moscow, Russia}

\date{\today} \maketitle

\begin{abstract}
Superfluid $^3$He-A and high-temperature superconductors both have gapless
fermionic quasiparticles
with the "relativistic" spectrum close to the gap nodes.  The interaction
of the "relaitivistic"
fermions with bosonic collective modes is described by the quantum field
theory, which results in a
close connection with particle physics. Many phenomena in high-energy
physics and cosmology can thus
be simulated in superfluid phases of $^3$He and in unconventional
superconductors. This includes
axial anomaly, vacuum polarization,  zero-charge effect, fermionic charge
of the vacuum,
baryogenesis,  event horizon,  vacuum instability,  Hawking  radiation,
etc. Analogs of some of
these phenomena, which are related to the axial anomaly, have been
experimentally simulated
in superfluid $^3$He. This includes  the  baryogenesis by textures
(Manchester), the
baryogenesis by cosmic strings (Manchester) and the generation of the
primordial magnetic field via the axial anomaly (Helsinki).
\end{abstract}
\
]

It is now well realized that the Universe and its broken symmetry ground state
-- the physical vacuum -- may behave like a condensed matter system, in which
the complicated degenerate ground state is developed due to the
symmetry breaking. The condensed matter counterparts of Universe
are represented by superconductors and by superfluid phases of $^3$He: both
systems contain quantum Bose and Fermi fields, describing the
interaction of the elementary quasiparticles with each other and with the
degenerate vacuum \cite{VolovikVachaspati,exotic}. This analogy allows us
to simulate in condensed
matter many properties of the physical vacuum, while the direct experiments
with the
physical vacuum are still far from the realization.

\begin{figure}[!!!t]
\begin{center}
\leavevmode
\epsfig{file=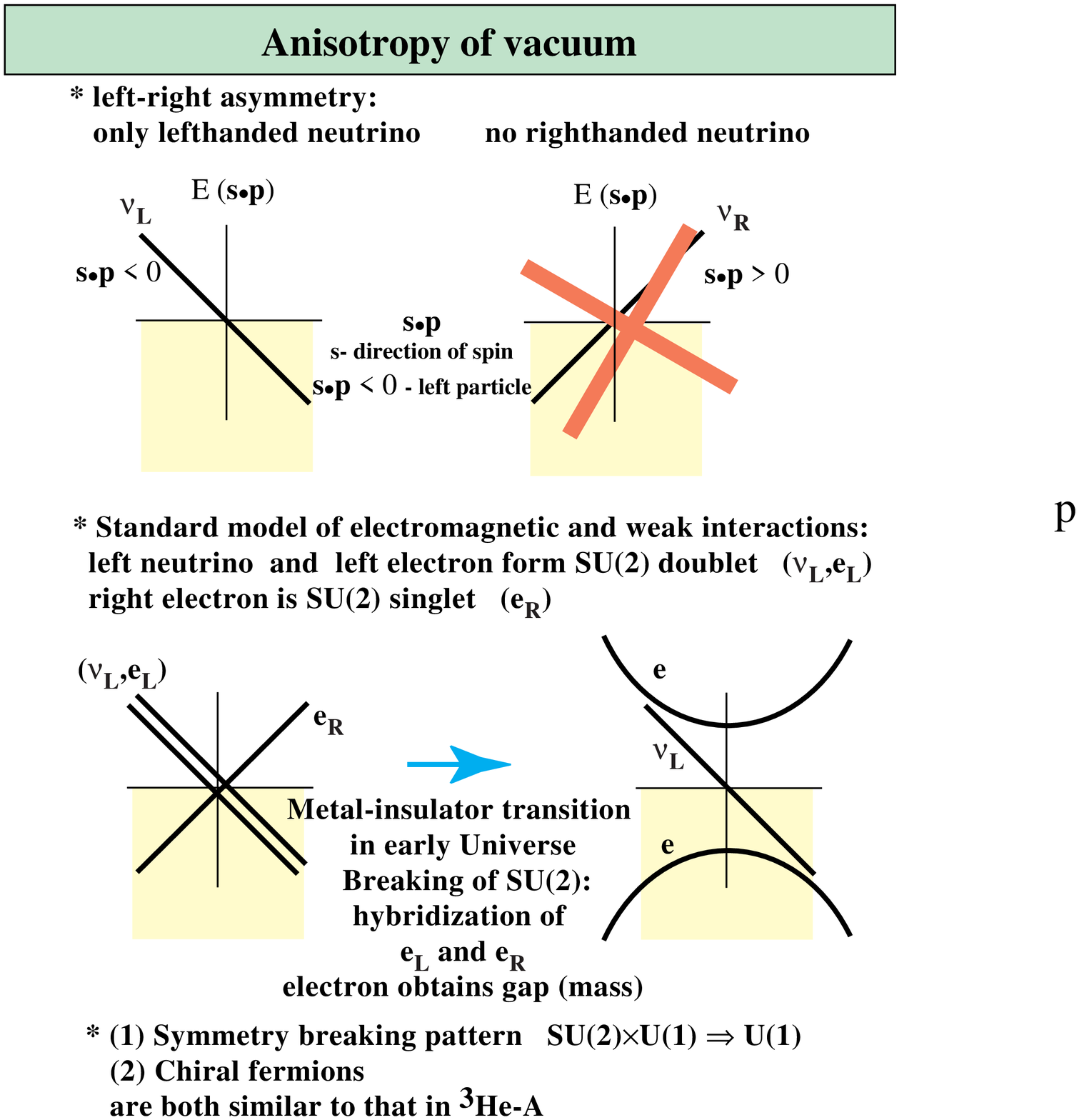,width=0.8\linewidth}
\caption[Metal-Insulator]
    {{\it (top)}: The spectrum of fermionic excitations of the physical
vacuum contain the branch of
the chiral particles: left-handed neutrino. The right-handed neutrino is
absent:  it is the
remarkable manifestation of the violation of the left-right symmetry in the
vacuum. Another
symmetry, which is broken in the present Universe, is the $SU(2)$ symmetry
of weak interactions.
In the unbroken symmetry state of the early Universe, the left
leptons (netrino and left electron) formed the
$SU(2)$ doublet, while the right electron was the $SU(2)$ singlet. {\it
(bottom)}: During the cool
down of the Universe the phase transition (or the crossover) occurred, at
which the $SU(2)$
symmetry was broken. As a result the left and right electrons were
hybridized forming the present
electronic spectrum with the gap
$\Delta=m_ec^2$. The electric properties of the vacuum thus exhibited the
metal-insulator
transition: The metallic state of the vacuum with the Fermi point  in the
elecronic spectrum was
transformed to the insulating state with the gap.}
\label{Metal-Insulator}
\end{center}
\end{figure}

The important property of the physical vacuum is its high anisotropy
(Fig.~\ref{Metal-Insulator}).

\begin{figure}[!!!!t]
\begin{center}
\leavevmode
\epsfig{file=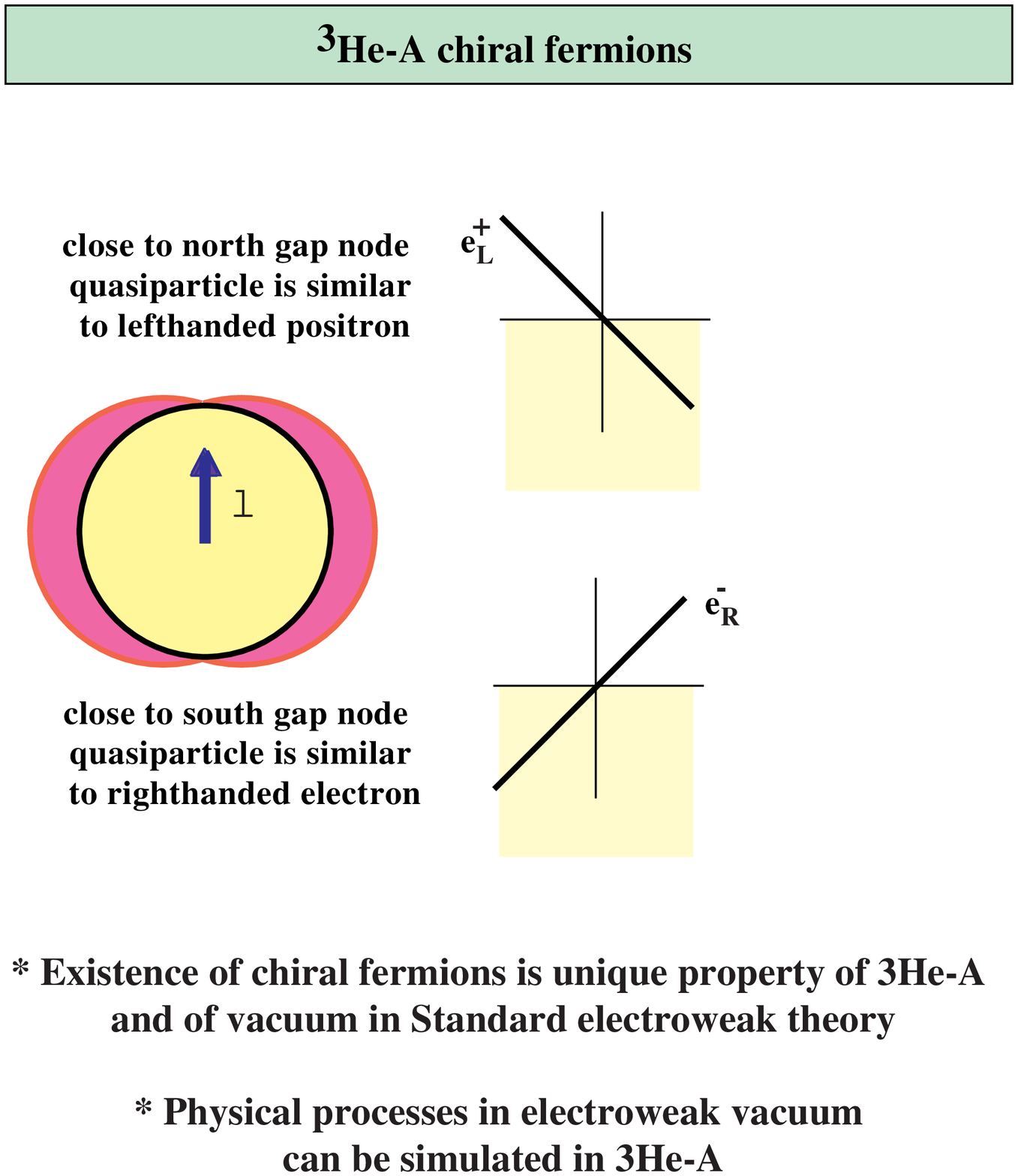,width=0.8\linewidth}
\caption[ChiralFermions]
   { Gap nodes and chiral fermions in $^3$He-A. Close to the gap nodes
the gapless quasiparticle represent the massless fermions moving
in the electromagnetic and gravity fields and obeying the relativistic equation
$g^{\mu\nu}(p_\mu - e A_\mu)(p_\nu - e A_\nu)=0$. Here $e=\pm$ is the
"electric charge" and simultaneously the chirality of the quasiparticles. The
righthanded particles have positive charge
$e=+1$ and are in the vicinity of the north pole, while the left particles with
$e=-1$ are in the vicinity of the south pole. $p_\mu=(E,{\bf p})$ is the
4-momentum;
$A_0=\mu_R$ plays the part of the chemical potential for the rifgt-handed
particles. The texture of  ${\hat{\bf l}}$ field provides the vector potential
${\bf A}=k_F{\hat{\bf l}}$ and also the metric tensor of the gravitational
field:
$g^{ik}= c_\perp^2 (\delta^{ik} - \hat l^i \hat l^k) +  c_\parallel^2 \hat l^i
\hat l^k ~~,~~ g^{00}=-1$, the quantities $c_\parallel= p_F/m$ and
$c_\perp= \Delta_0/p_F$ correspond to velocities of "light" propagating
along and transverse to  ${\hat{\bf l}}$, here $m$ is the mass of the $^3$He
atom and
$\Delta_0$ is the amplitude of the gap.}
\label{ChiralFermions}
\end{center}
\end{figure}

\begin{figure}[!!!t]
\begin{center}
\leavevmode
\epsfig{file=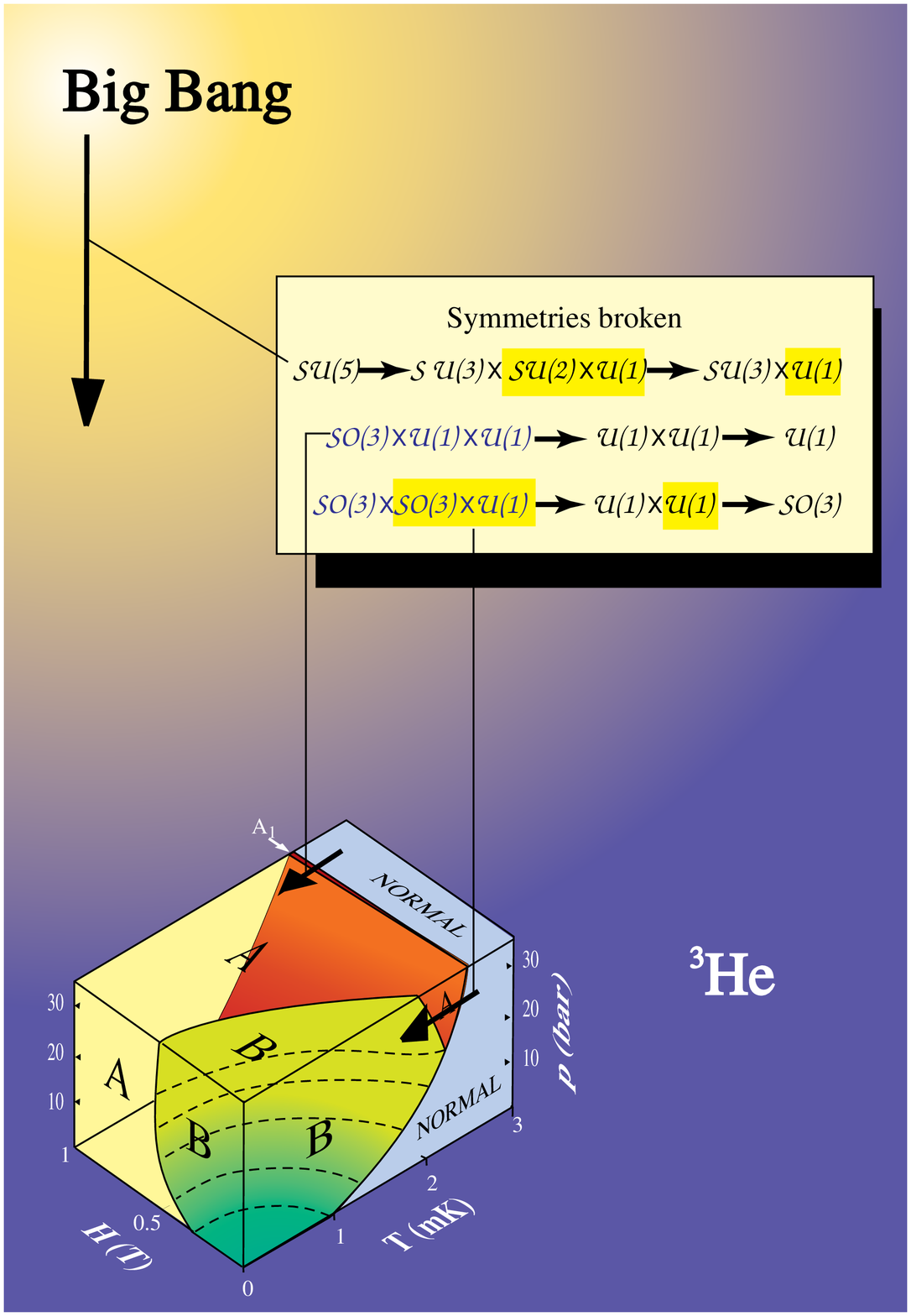,width=0.8\linewidth}
\caption[BigBangVs3He]
    {The symmetry breaking in the electroweak transitions is similar to
that in the transition
from the normal $^3$He to $^3$He-A.}
\label{BigBangVs3He}
\end{center}
\end{figure}

The physical vacuum (superfluid ground state) of superfluid $^3$He-A
resembles the electroweak
vacuum. The fermionic excitations of the $^3$He-A in the vicinity of
the gap nodes are "relativistic" chiral fermions
(Fig.~\ref{ChiralFermions}). They are the
left-handed near the north pole, where they have a linear momentum ${\bf
p}$ close to $p_F{\hat{\bf
l}}$ (${\hat{\bf l}}$ is the direction of spontaneous angular momentum in
$^3$He-A) and they are
right-handed near the south pole where ${\bf p}=-p_F{\hat{\bf l}}$. Another
important similarity is
that the symmetry breaking pattern   $SU(2)\times U(1) \rightarrow U(1)$ in
electro-weak
interactions is the same as the symmetry
breaking pattern  $SO(3)\times U(1)\rightarrow U(1) $  in $^3$He-A
(Fig.~\ref{BigBangVs3He}).

\begin{figure}[!!!t]
\begin{center}
\leavevmode
\epsfig{file=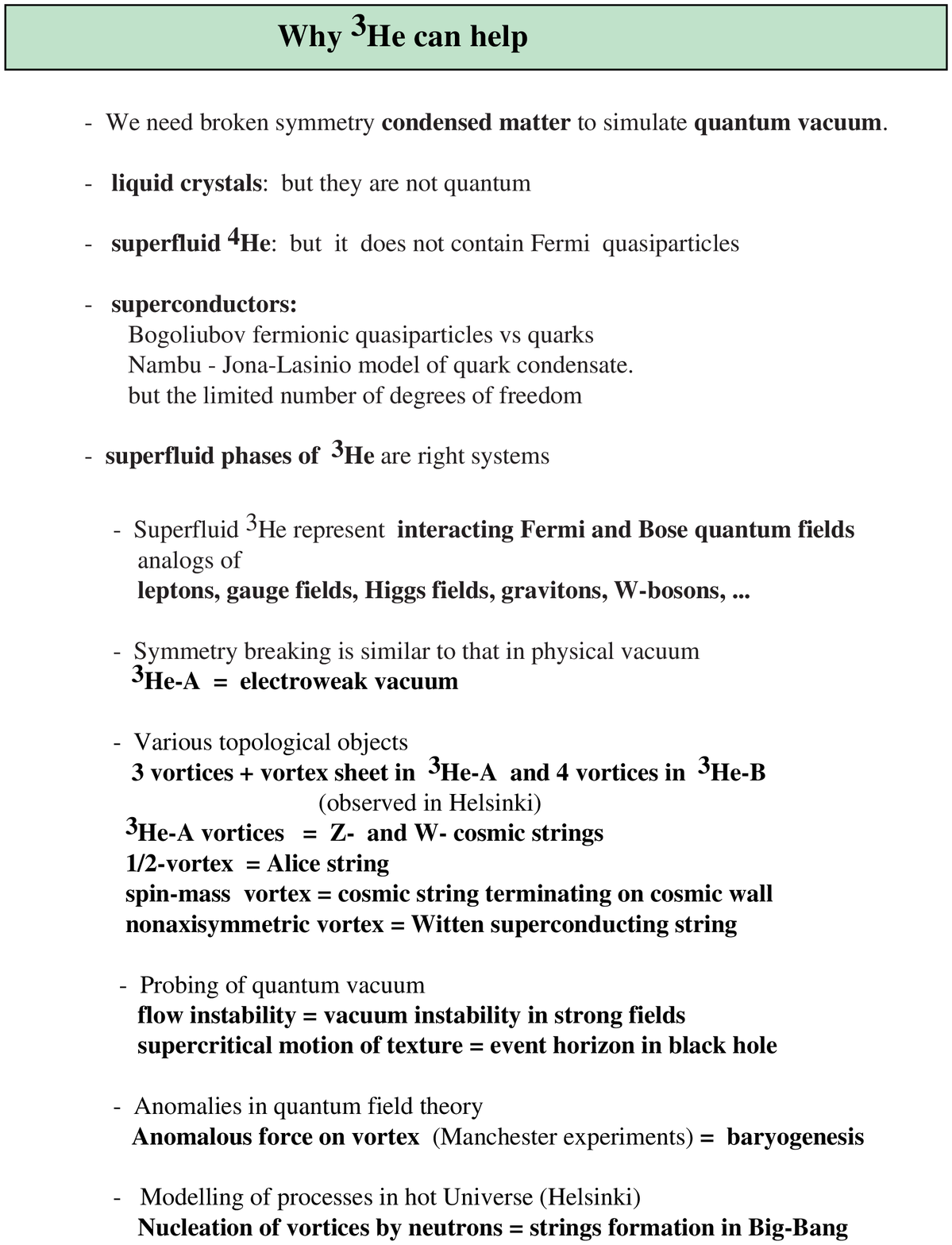,width=0.8\linewidth}
\caption[3HeHelp]
    {Superfluid $^3$He is the right condensed matter for simulation of the
properties of the
physical vacuum.}
\label{3HeHelp}
\end{center}
\end{figure}

\begin{figure}[!!!t]
\begin{center}
\leavevmode
\epsfig{file=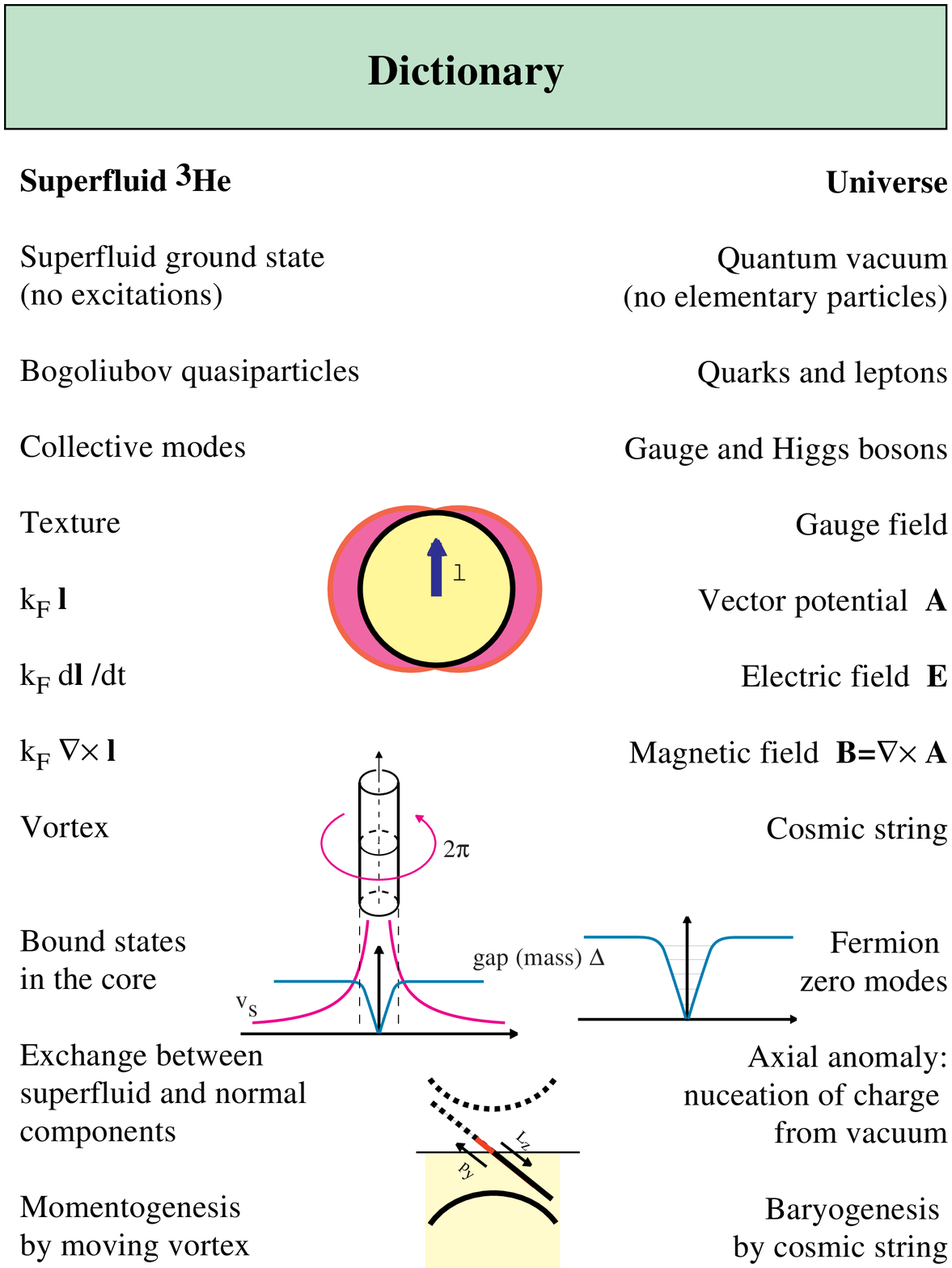,width=0.8\linewidth}
\caption[Dictionary]
    {Dictionary}
\label{Dictionary}
\end{center}
\end{figure}

That is why the physical effects are actually the same in the vacuum of the
high energy physics and
in superfluid $^3$He-A. Thus the $^3$He-A is the right condensed matter for
simulations of the vacuum
properties (Fig.~\ref{3HeHelp}).
The main difference is in the terminology, so we need the
dictionary for the translation from the particles physics language to that
of $^3$He-A
(Fig.~\ref{Dictionary}).

\begin{figure}[!!!t]
\begin{center}
\leavevmode
\epsfig{file=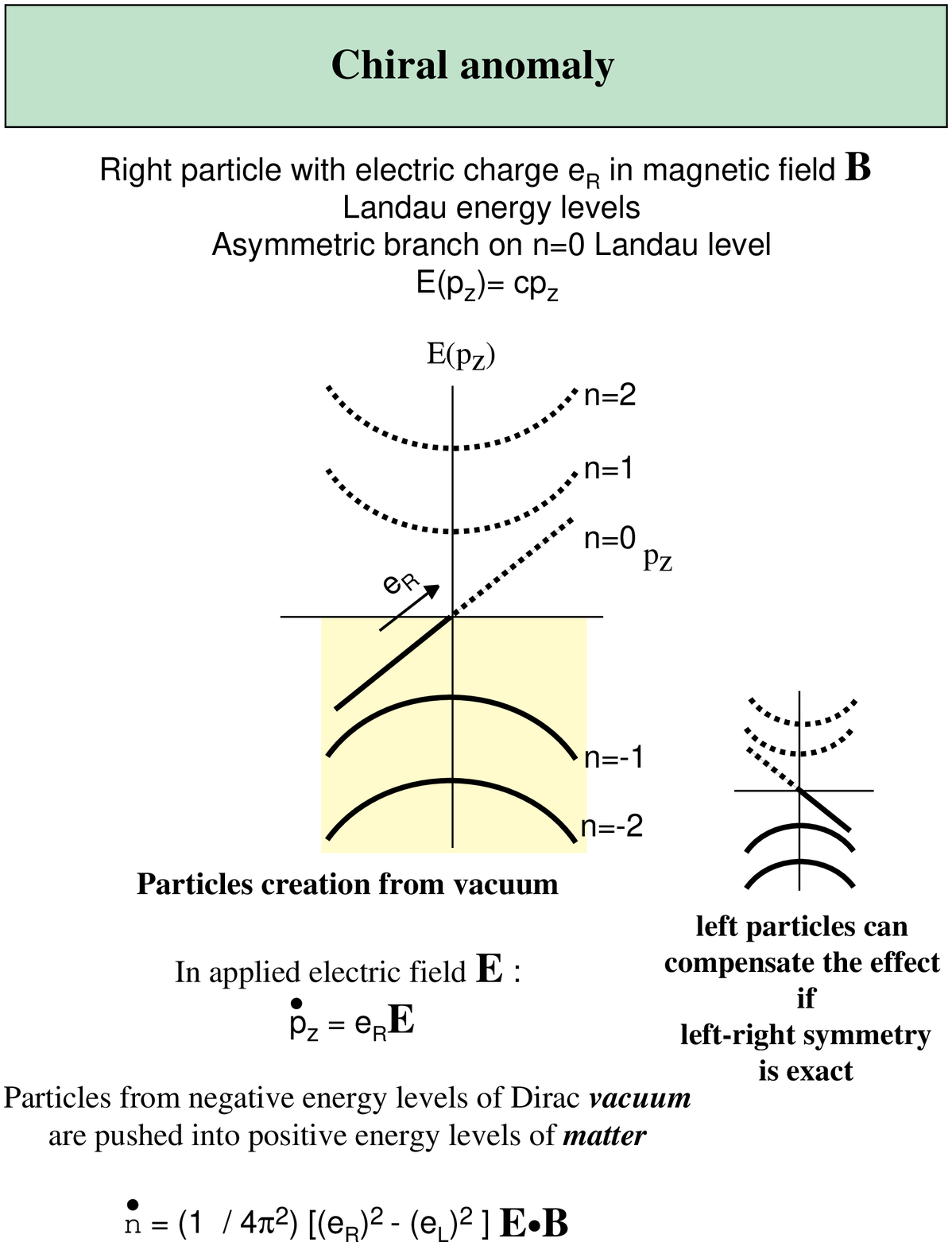,width=0.8\linewidth}
\caption[ChiralAnomaly]
   {Chiral anomaly: the spectral flow from the occupied energy levels caused
by the applied electric field leads to the creation of the fermionic charge
from the vacuum if the
left-right symmetry is violated. }
\label{ChiralAnomaly}
\end{center}
\end{figure}

\begin{figure}[!!!t]
\begin{center}
\leavevmode
\epsfig{file=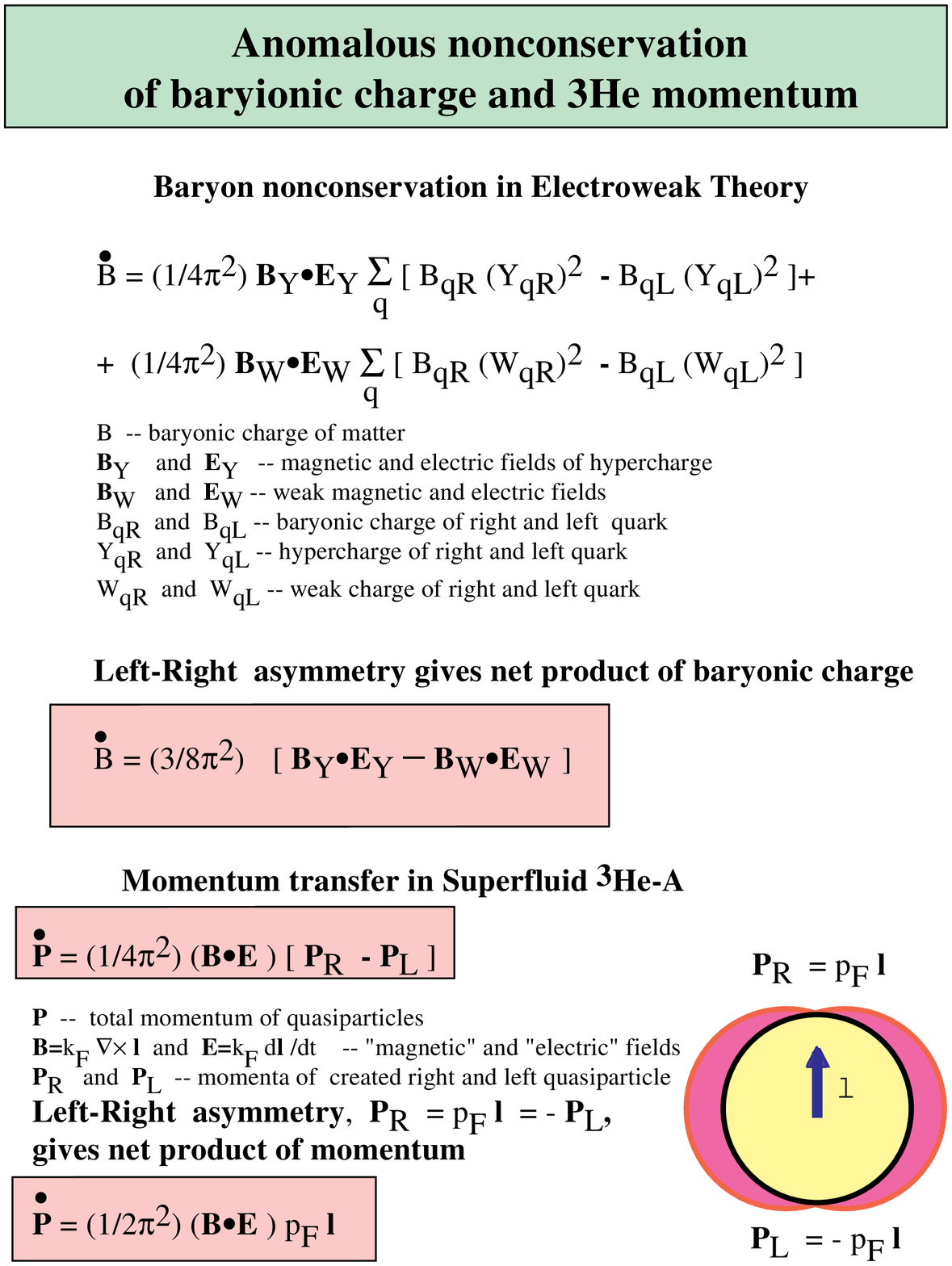,width=0.8\linewidth}
\caption[AnomalousNonconservation]
    {The same anomaly equation leads to production of the baryonic charge
$B$ in high energy physice
and to production of the  linear momentum ${\bf P}$ in superfluid
$^3$He-A.}
\label{AnomalousNonconservation}
\end{center}
\end{figure}

\begin{figure}[!!!t]
\begin{center}
\leavevmode
\epsfig{file=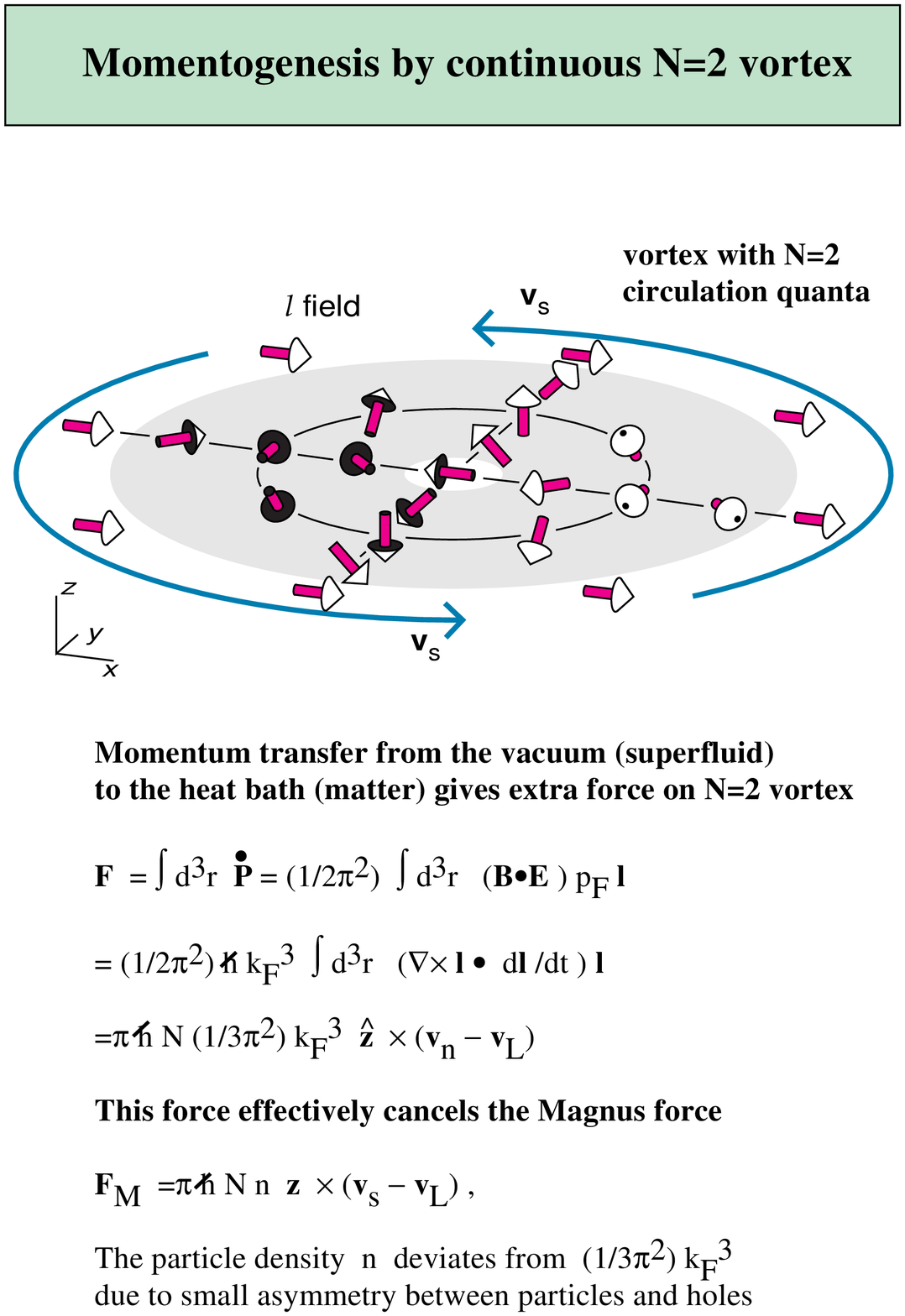,width=0.8\linewidth}
\caption[ContinuousMomentogenesis]
    {The order
parameter texture (analog of magnetic field) in the core of continuous
vortex  is
converted into the excess of the fermionic charge (linear momentum) due to
the axial anomaly in the moving continuous vortex. The moving vortex
generates the time dependence of
the order parameter, which is equivalent to electric field. }
\label{ContinuousMomentogenesis}
\end{center}
\end{figure}

\begin{figure}[!!!t]
\begin{center}
\leavevmode
\epsfig{file=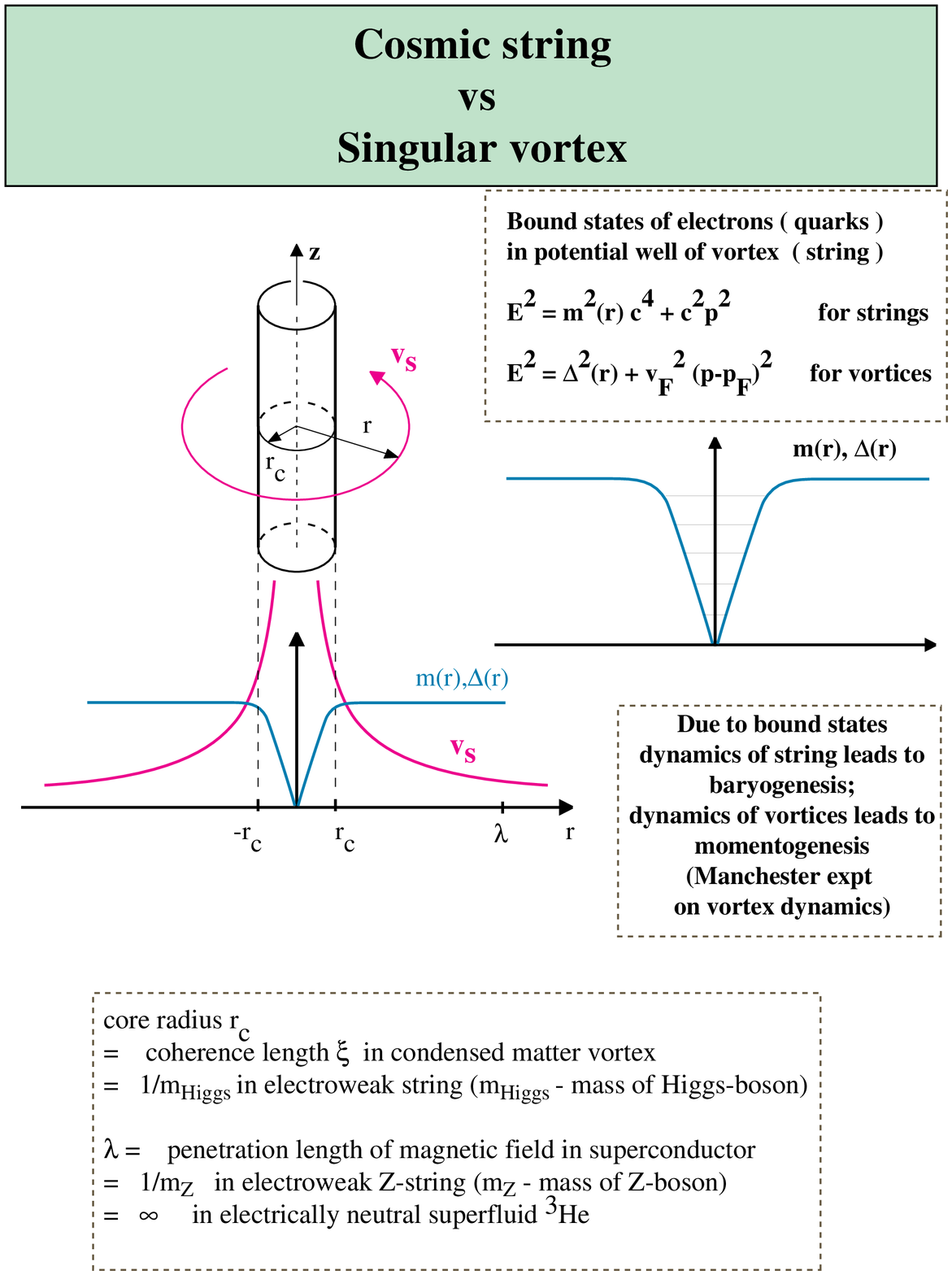,width=0.8\linewidth}
\caption[VortexString]
    {Cosmic string is the counterpart of the Abrikosov vortex in
superconductor. It also contains
the bound states of the fermions, which are important for the baryoproduction.}
\label{VortexString}
\end{center}
\end{figure}

\begin{figure}[!!!t]
\begin{center}
\leavevmode
\epsfig{file=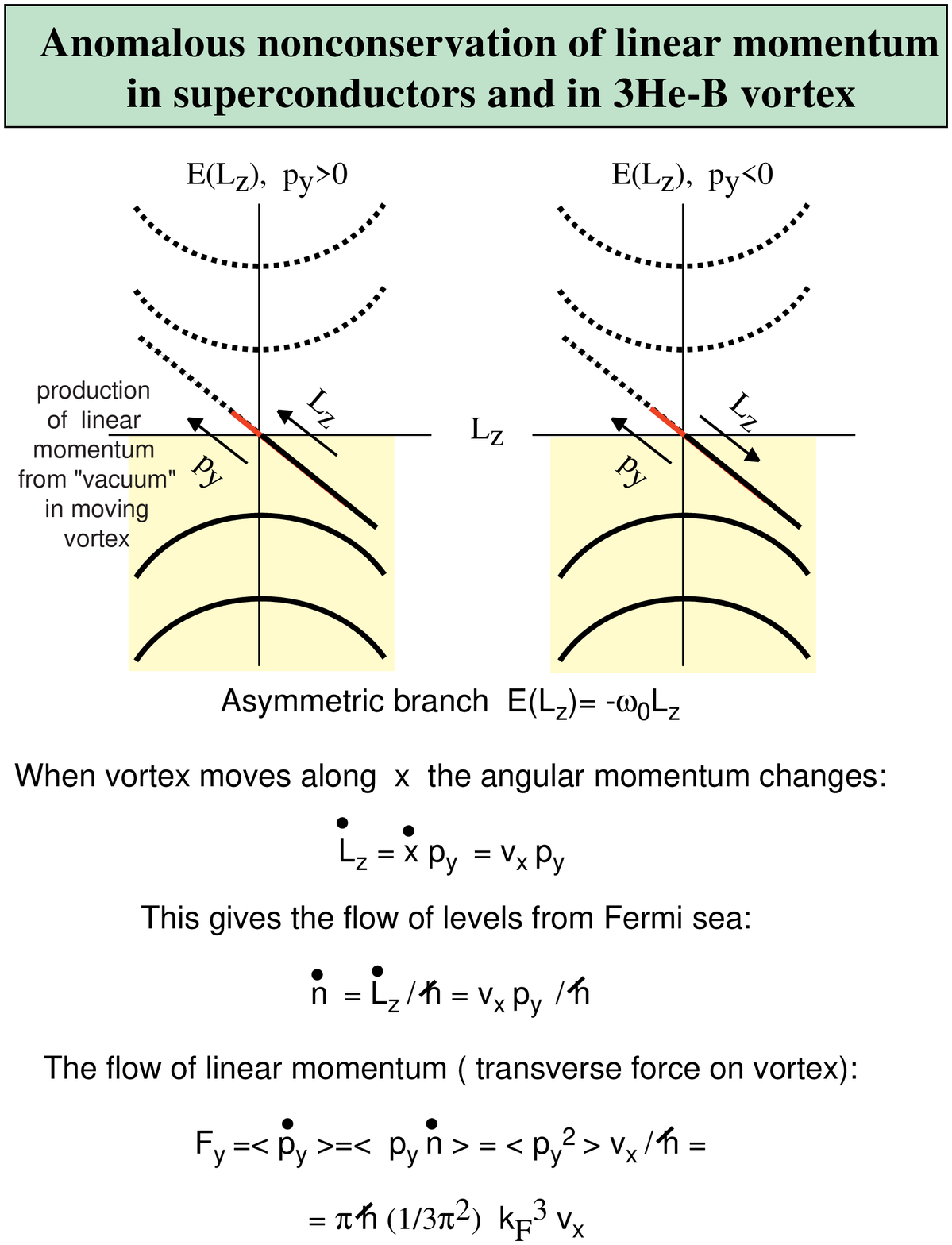,width=0.8\linewidth}
\caption[SingularMomentogenesis]
    {The spectral flow of the momentum in the core of the moving singular
vortex leads to the
experimentally observed reactive force on a vortex in superfluid $^3$He-B.}
\label{SingularMomentogenesis}
\end{center}
\end{figure}

\begin{figure}[!!!t]
\begin{center}
\leavevmode
\epsfig{file=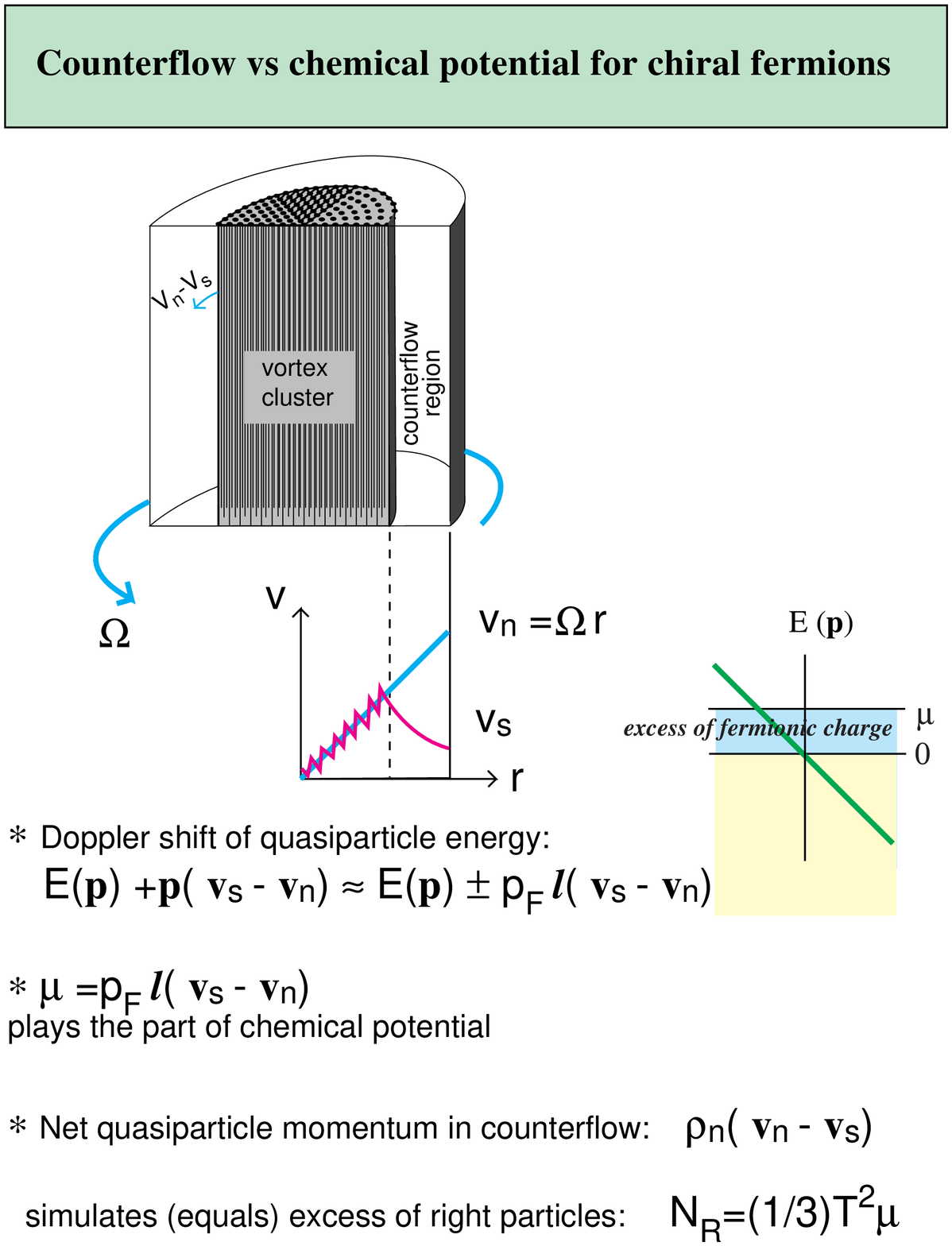,width=0.8\linewidth}
\caption[Counterflow]
    {Instability of the counterflow.  The counterflow generated by the
rotation of the cryostat is
equivalent to the nonzero chemical potential for the right electrons. When
the velocity of the
counterflow
${\bf v}_n - {\bf
v}_s$ in the ${\hat{\bf l}}$ direction (corresponding to the chemical potential
$\mu_R$ of the chiral electrons)   exceeds some critical value, the instability
occurs and the container  becomes filled with the ${\hat{\bf l}}$-texture which
corresponds to the hypermagnetic field.}
\label{Counterflow}
\end{center}
\end{figure}

\begin{figure}[!!!t]
\begin{center}
\leavevmode
\epsfig{file=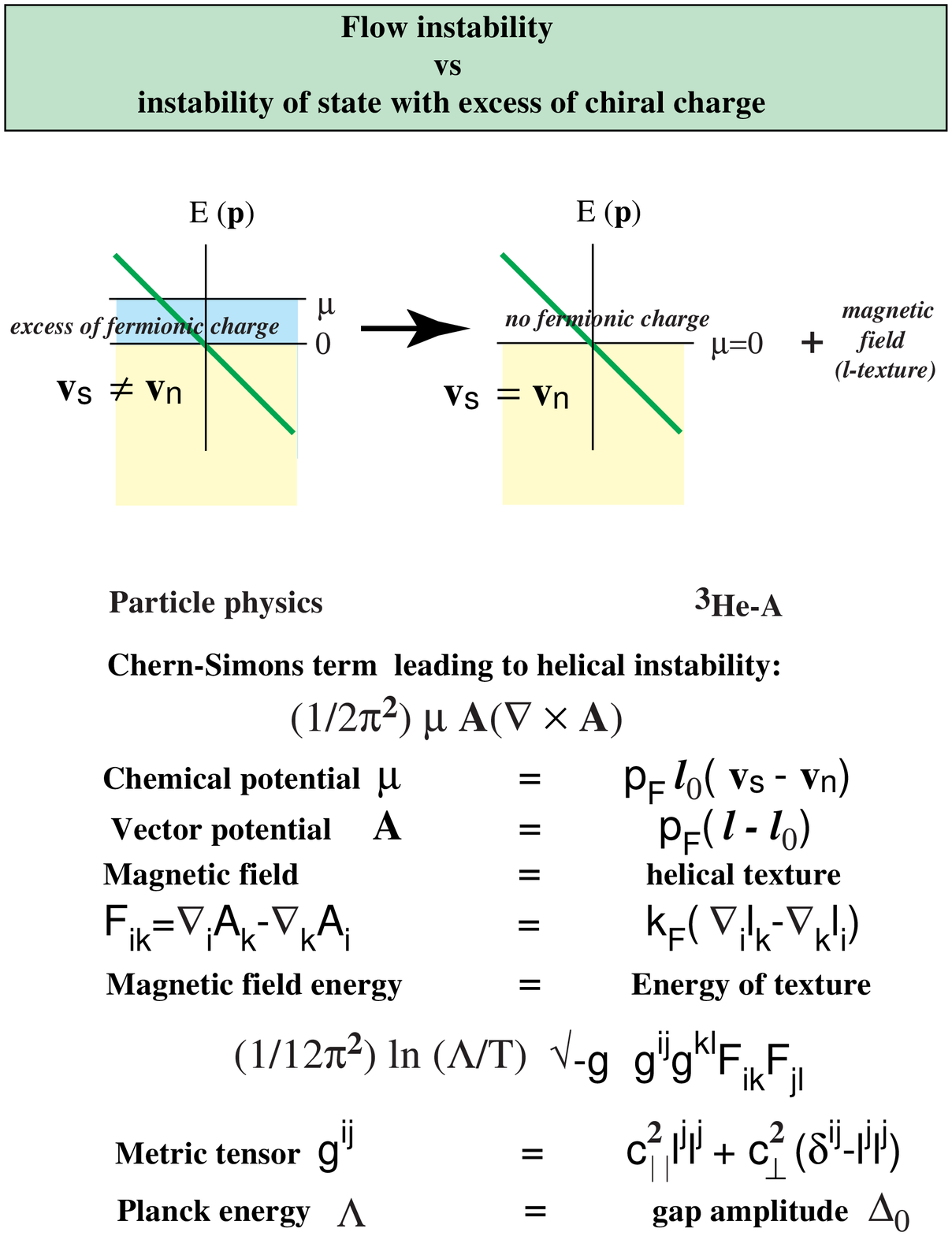,width=0.8\linewidth}
\caption[PrimordialField]
    {The excess of the
chiral right-handed electrons in early Universe can be effectively
converted to the hypermagnetic field due to the mechanism of chiral
anomaly.  This is described
essentially by the same equations as the counterflow instability observed
in $^3$He-A.}
\label{PrimordialField}
\end{center}
\end{figure}

Here we discuss 3 experiments in superfluid $^3$He which simulate the
processes in the early
Universe. The first two are related to the cosmological problem why the
Universe contains much more
matter than antimatter. The $^3$He experiments allow  to test different
scenaria in which the
nonconservation of the baryons is caused by the so called axial anomaly,
which governs  the
nucleation of the fermionic charge from the vacuum.  The essence of the
axial anomaly is presented
in Fig.~\ref{ChiralAnomaly}. The chiral anomaly equation
\cite{Adler1969,BellJackiw1969} is
obtained using the picture of  Landau levels for the right-handed particle
in the applied magnetic
field.
This anomaly equation which describes the nucleation of the fermionic
charges in the presence of
magnetic and electric fields describes both the production of the baryons
in the
electroweak vacuum (baryogenesis) and  the  production of  the  linear
momentum in the
superfluid
$^3$He-A (momentogenesis) (Fig.~\ref{AnomalousNonconservation}). In
$^3$He-A the effective magnetic
and electric fields are generated  by the moving
continuos vortex texture. The anomalous production of the linear momentum
leads to
the additional force acting on the continuous vortex in $^3$He-A
(Fig.~\ref{ContinuousMomentogenesis}), which was measured in the Manchester
experiment
\cite{BevanNature}.  The anomaly equation has thus been verified in the
$^3$He-A experiments with
the moving vortices.
The axial anomaly is also responsible for the dynamics of singular vortices
and for the
baryogenesis by the singular cosmic string. In this
case the spectral flow from the vacuum is realized by the bound states of
quasiparticles in the core of vortices and strings
(Figs.~\ref{VortexString},\ref{SingularMomentogenesis}).  The
baryoproduction in the core of the
cosmic string was simulated in the Manchester experiment on dynamics of
singular vortices in
$^3$He-B  \cite{BevanNature}.
There is also an opposite effect:  the anomalous transformation of the
chiral particles into the magnetic field. In the Universe this mechanism
can be responsible for
the formation of the primordial magnetic field \cite{JoyceShaposhnikov},
while in the $^3$He-A the
similar anomaly equations describes the collapse of the excitation momenta
towards the formation of the textures -- counterpart of the magnetic field
in cosmology (Figs.~\ref{Counterflow},\ref{PrimordialField}).  Such
collapse was recently observed
in  Helsinki rotating cryostat \cite{Experiment}.

\begin{figure}[!!!t]
\begin{center}
\leavevmode
\epsfig{file=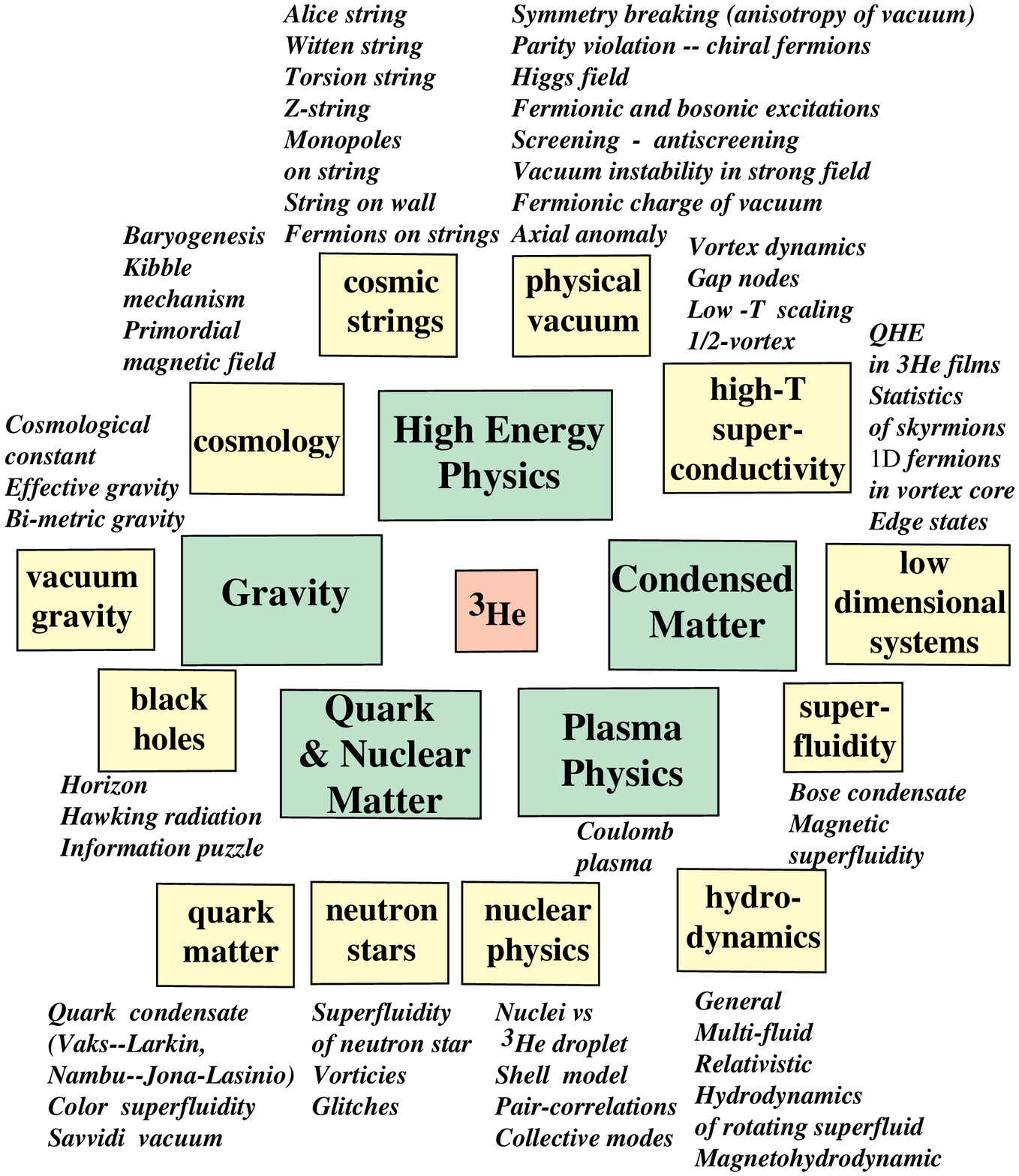,width=0.8\linewidth}
\caption[Connections]
    {Connections of  superfluid $^3$He to other branches of physics.}
\label{Connections}
\end{center}
\end{figure}

We discussed only 3 experiments in superfluid $^3$He related to the
properites of the electorweak
vacuum. In all of them the chiral
anomaly is an important mechanism. It regulates the nucleation of the
fermionic charge from the
vacuum as was observed in \cite{BevanNature} and the inverse process of the
nucleation of the
effective magnetic field from the fermion current as was observed in
\cite{Experiment}. There are
many other connections between the superfluid $^3$He and other branches of
physics which should be
exploited (Fig.~\ref{Connections}).

\vfill\eject

\end{document}